\documentclass[useAMS,usenatbib]{mn2e}
\usepackage{lscape}
\usepackage{graphicx,times}
\usepackage{float}
\usepackage{natbib}
\title[Multiwavelength study of Cygnus A III. Evidence for relic lobe plasma]{Multiwavelength study of Cygnus~A III. Evidence for relic lobe plasma}
\author[K.C. Steenbrugge, I. Heywood and K. M. Blundell]{Katrien
  C. Steenbrugge$^{1}$\thanks{E-mail:kcs@astro.ox.ac.uk}, Ian
  Heywood$^{2}$ and Katherine M. Blundell$^{2}$ \\ 
  $^{1}$St John's
  College Research Centre, University of Oxford, St John's College,
  Oxford, OX1 3JP, UK \\ 
$^{2}$University of Oxford, Department of
  Physics, Keble Road, Oxford, OX1 3RH, UK}

\begin{document}

\date{Accepted . Received }

\pagerange{\pageref{firstpage}--\pageref{lastpage}} \pubyear{2002}

\maketitle

\label{firstpage}

\begin{abstract}

We study the particle energy distribution in the cocoon surrounding
Cygnus A, using radio images between 151~MHz and 15~GHz and a 200~ks {\it
Chandra} ACIS-I image. We show that the excess low frequency emission
in the the lobe further from Earth cannot be explained by absorption
or excess adiabatic expansion of the lobe or a combination of
both. We show that this excess emission is consistent with emission
from a relic counterlobe and a relic counterjet that are being
re-energized by compression from the current lobe. We detect hints of a relic
hotspot at the end of the relic X-ray jet in the more distant
lobe. We do not detect relic emission in the lobe nearer to Earth as
expected from light travel-time effects assuming intrinsic symmetry. We
determine that the duration of the previous jet activity phase was
slightly less than that of the current jet-active phase. Further, we
explain some features observed at 5 and 15~GHz as due to the presence
of a relic jet.  
\end{abstract}


\section{Introduction\label{sect:intro}}

At a redshift of 0.05607 \citep{owen97}, Cygnus~A
(3C~405) is the
closest powerful FR~II classical double radio galaxy. It is therefore
well studied in the radio \citep{carilli91}. At this redshift, and assuming
a Hubble constant of 73 km s$^{-1}$ Mpc$^{-1}$, an angular size of 1 arcsecond
corresponds to a physical size of 1.044~kpc. In this paper, as
in the previous two in this series, we use `lobe' to mean the lobe that is
nearer to Earth, and `counterlobe' for its more distant
counterpart. Similarly, `jet' refers to the jet pointing towards the
Earth, while `counterjet' is used for the receding jet.
 
In a previous paper we showed that the current radio jet axis of
Cygnus~A appears to be precessing \citep{steenbrugge08a} and derived
an upper limit to the jet speed of 0.5$c$. In a companion paper
\citep{steenbrugge08b} we demonstrated that the long narrow X-ray
feature observed in the counterlobe is a relic counterjet. This
conclusion stems from the fact that the outer part of this linear
feature does not trace the current counterjet observed in the 5, 8 and
15 GHz images. The relic counterjet is wider than the current radio
jets, and emits X-rays via inverse-Compton scattering of the cosmic
microwave background, rather than synchrotron as the current jet and
counterjet.

The age of the current jet-activity is only of order 4$\times$
10$^7$ years, assuming a constant hotspot advance speed of
0.005c \citep{alexander96}. Furthermore, assuming a jet speed of
0.3$c$ \citep{steenbrugge08a}, the time for an emitted plasma blob to
reach the hotspot is 7$\times$10$^5$ years.

In this paper, we explore the possibility that relic radio lobes, formed by
the relic (counter)jet, may be observable. The reigning paradigm in
this field, which dates back to early classic works by
e.g. \citet{longair73} and \citet{scheuer74}, is that intrinsically
both lobes have similar properties. \citet{dennett-thorpe99} studying
nearby FR II galaxies found no correlation between jet-side and
spectral index for the extended lobe emission, and showed that the
hotspot spectra on both sides are similar. Both these observations
support the assumption that intrinsically lobes formed by symmetric
and active jets have the same properties. We investigate the
possibility of relic lobe emission using the following methods: i) we
compare the total low frequency flux in the lobe and counterlobe, ii)
the same as i) but for portions of the (counter)lobe and iii) by
studying the morphology at high frequencies. For this purpose we
study the low frequency (151, 327 and 1345~MHz), yet high resolution
data available for Cygnus~A. We compare these low frequency data with
the high frequency 5-GHz radio image and with the 200~ks {\it Chandra}
ACIS-I image of Cygnus~A. The 151~MHz, 327~MHz, 1345~MHz and X-ray
images of Cygnus~A show significantly more emission closer to the
nucleus than the higher frequency (5 $-$ 15~GHz) radio images (see
Figs.~\ref{fig:volume} and \ref{fig:relic}), as noted previously
by \cite{carilli91} and \cite{lazio06}. The X-ray emission close to
the nucleus is found to be thermal \citep{wilson06}. The radio
emission at the same location however, in common with radio emission
throughout this source, is synchrotron emission.

In classical double FRII radio sources, of which Cygnus~A is the
prototype, we contend that  adiabatic expansion is the dominant
mechanism by which the electrons lose energy with time and emit at lower
frequencies. The generality of this effect is supported by the
D-$\alpha$ correlation obtained by \cite{blundell99} and explored more
generally in \citet{blundell00,blundell01}. \cite{rudnick94} 
reported that they found no evidence, in a detailed pixel-by-pixel
study of Carilli's data on Cygnus A, of any aging of the spectrum due
to synchrotron losses or inverse-Compton losses, consistent with the
hypothesis that E$^2$ losses are negligible. An alternative picture,
critiqued by \cite{blundell01}, is that synchrotron and inverse Compton
losses dominate the evolution of the electron energy spectrum in the
lobes; this picture underpins the traditional spectral ageing
method. The assumptions which underlie this picture are not
necessarily applicable to FRII sources such as Cygnus A
\citep{blundell00,blundell01}.
Thus low-frequency emission
may represent aged electrons, relative to those radiating at GHz
frequencies. In this paper we study whether the relic X-ray detected
counterjet \citep{steenbrugge08b} has observable effects on the
counterlobe radio emission. In particular, we test the hypothesis that
relic counterlobe plasma is detectable at low radio frequencies.

\section{Data reduction and cleaning}

The 151-MHz MERLIN image was kindly provided by Paddy Leahy and
published by \cite{leahy89}. 

The 327~MHz VLA\footnote{ The Very Large Array is a facility of the
National Radio Astronomy Observatory, National Science Foundation.},
the 1345~MHz, the 5~GHz, the 8~GHz and 15~GHz data were kindly supplied by
Chris Carilli; and published by \cite{carilli91}, \cite{carilli96},
\cite{carilli96b} and \cite{perley96}. 

The {\it Chandra} image presented in this paper is the co-added image
of 200~ks of ACIS-I data. The X-ray data reduction and co-adding of
the different images is described by \cite{steenbrugge08b}. The X-ray
spectra were fitted using the SPEX code \citep{kaastra02} and the
errors for the X-ray parameters are rms errors
\citep{kaastra04}. However, for the spectral result we used all 10
{\it Chandra} datasets with more than 5~ks exposure time \citep[for
more details see][]{steenbrugge08b}. All the spectra were fitted
simultaneously. We allowed for a fudge factor for the normalisation of
the power-law component to take into account the different calibration
between the ACIS-S and ACIS-I spectra.

\section{Radio spectra and fluxes}

The 5-GHz and higher frequency images of Cygnus~A show a lack of
emission near the nucleus \citep[e.g.][]{steenbrugge08b}. The 5-GHz
multi-configuration VLA image obtained by \cite{carilli91} has the
highest dynamic range of all the high frequency radio images, and it
was therefore used as a comparison to the lower frequency data. To
ensure that no diffuse emission at 5~GHz is missed, we re-imaged the
radio data, and increased the weighting of the shorter baselines. This
did not result in any additional detected 5~GHz emission near the
nucleus. The upper limit to the flux density near the nucleus is
0.5~Jy over 1233.7 beams or 616.9 arcsec$^2$ at 5~GHz. Furthermore,
for the highest resolution image the total flux density at 5~GHz is
383 $\pm$ 9~Jy. This is similar to the flux density of 410.6~Jy as
predicted by the model of \citet{baars77}. We conclude that the lack
of emission at 5~GHz and higher frequencies near the nucleus is not an
instrumental or undersampling effect, but due to a lack of
sufficiently high Lorentz factor electrons.

\begin{figure}
\begin{center}
 \resizebox{\hsize}{!}{\includegraphics[angle=0]{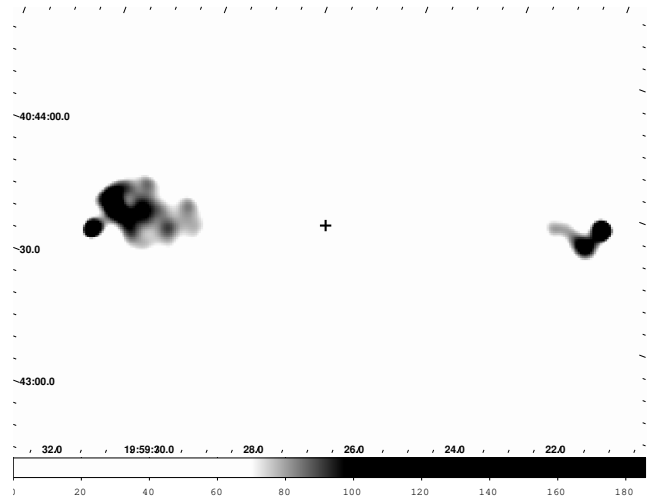}}
\caption{The 151~MHz image of Cygnus~A with a transfer function
  emphasising the excess flux coming from the counter (i.e. eastern)
  lobe. The cross indicates where the nucleus is detected in X-ray and
  higher radio frequencies. The grey scale indicates the flux
  density in Jy per beam. \label{fig:151}}
\end{center}
\end{figure}

By examining the 151 and 327~MHz radio images it becomes apparent
that the counterlobe is brighter than the lobe in the outermost region
(see Fig.~\ref{fig:151}). This is consistent with the results obtained
at 74~MHz by \cite{kassim96}. To see if this brightness difference
persists at higher frequencies, we determined the radio flux densities
for the lobe and counterlobe for the following bands: 151, 327 and
1345~MHz, and 5, 8 and 15~GHz. 

For each band we used the highest spatial resolution image available,
and only included emission above a certain threshold, as described
below. Rather than measuring the flux density in a single template
region common to all frequency bands the threshold varied with
frequency. This threshold value was determined by blanking all the
emission in the map below a starting threshold which was iteratively
reduced.  The total flux density for each map in the resulting series
was determined and compared to the value predicted by \citet{baars77}.
Adjacent pairs of maps were then differenced, which showed the extra
emission that was being added for each iteration. We selected the
threshold which was closest to the value predicted by the
\citet{baars77} model, but did not introduce either background noise
spikes in the map or features that we deemed to be due to residual
calibration errors. This was repeated for each frequency. 

To ensure a reliable measurement and to prevent overestimating the
flux density the iterations were terminated when spurious features
were introduced into the map (e.g. noise spikes or features that we
deemed were due to residual calibration errors). This was repeated for
each frequency. As an additional check we also compared the total
measured flux density to those from the model derived by \cite{baars77} and in no case do we exceed this value.

A slight complication is that the hotspot and jets have spectral
indices which differ from that of the lobe material and could
potentially contaminate the spectra determined for the lobe or the
counterlobe. Therefore the flux densities from both hotspots and
counter-hotspots were removed from the lobe and counterlobe flux
densities. We used the AIPS task JMFIT to fit each hotspot with
a Gaussian component and a two-dimensional quadratic background
surface. The Gaussian component was then subtracted from the
(counter)lobe for each hotspot at each frequency.


At frequencies above 327~MHz there are two hotspots detected in each
lobe, however at 151 and 327~MHz only the brightest hotspot is
detected in the counterlobe. At any given frequency we only measure
and subtract the contributions from hotspots which are detected, and
do not attempt to compensate for the potential contribution to the
flux density measurements from hotspots which are not detected. 

\begin{figure}
\begin{center}
\begin{minipage}{0.5\textwidth}
\includegraphics[width=0.95\textwidth,clip=t]{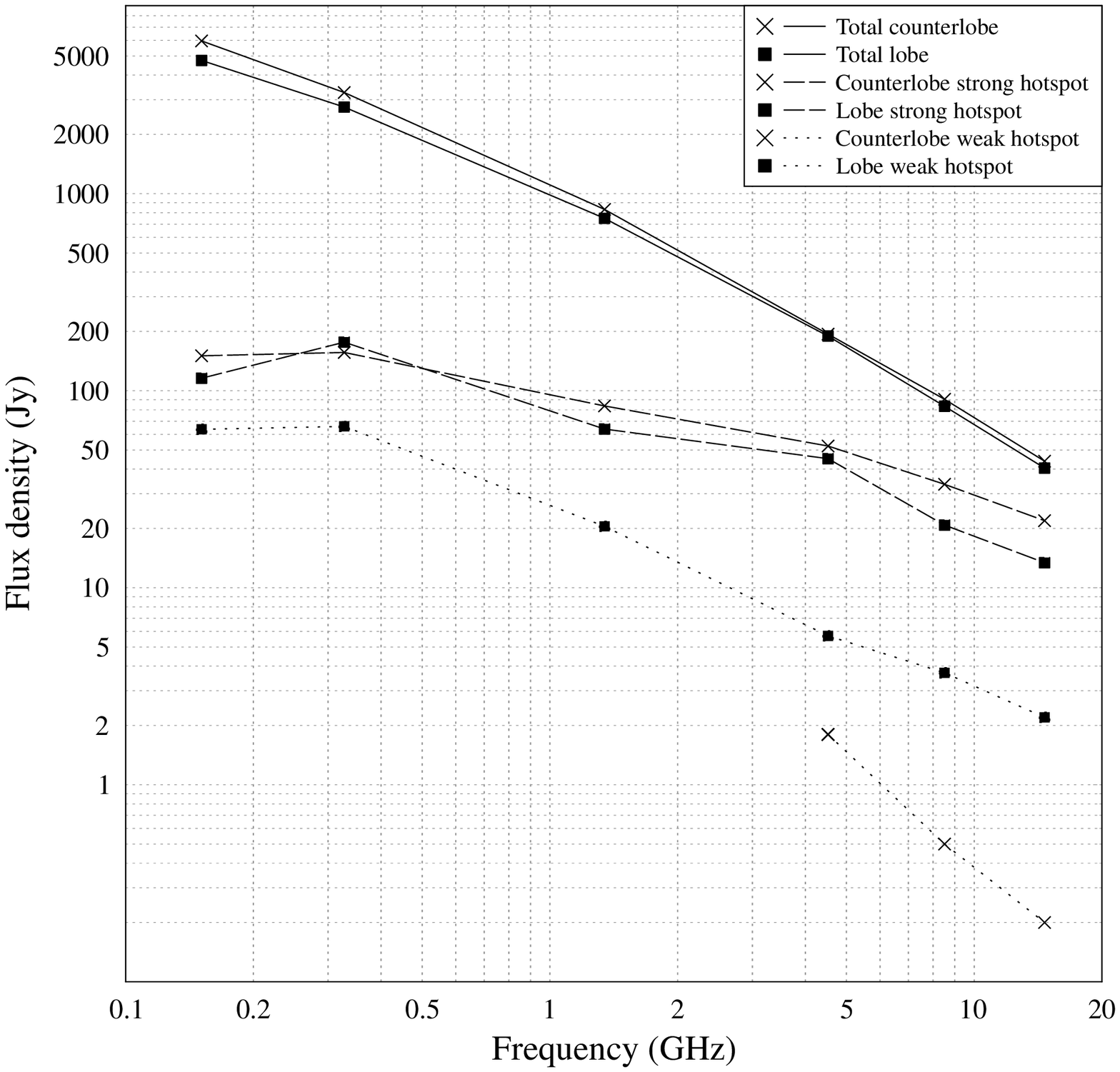}
\end{minipage}
\begin{minipage}{0.5\textwidth}
\includegraphics[width=0.95\textwidth,clip=t]{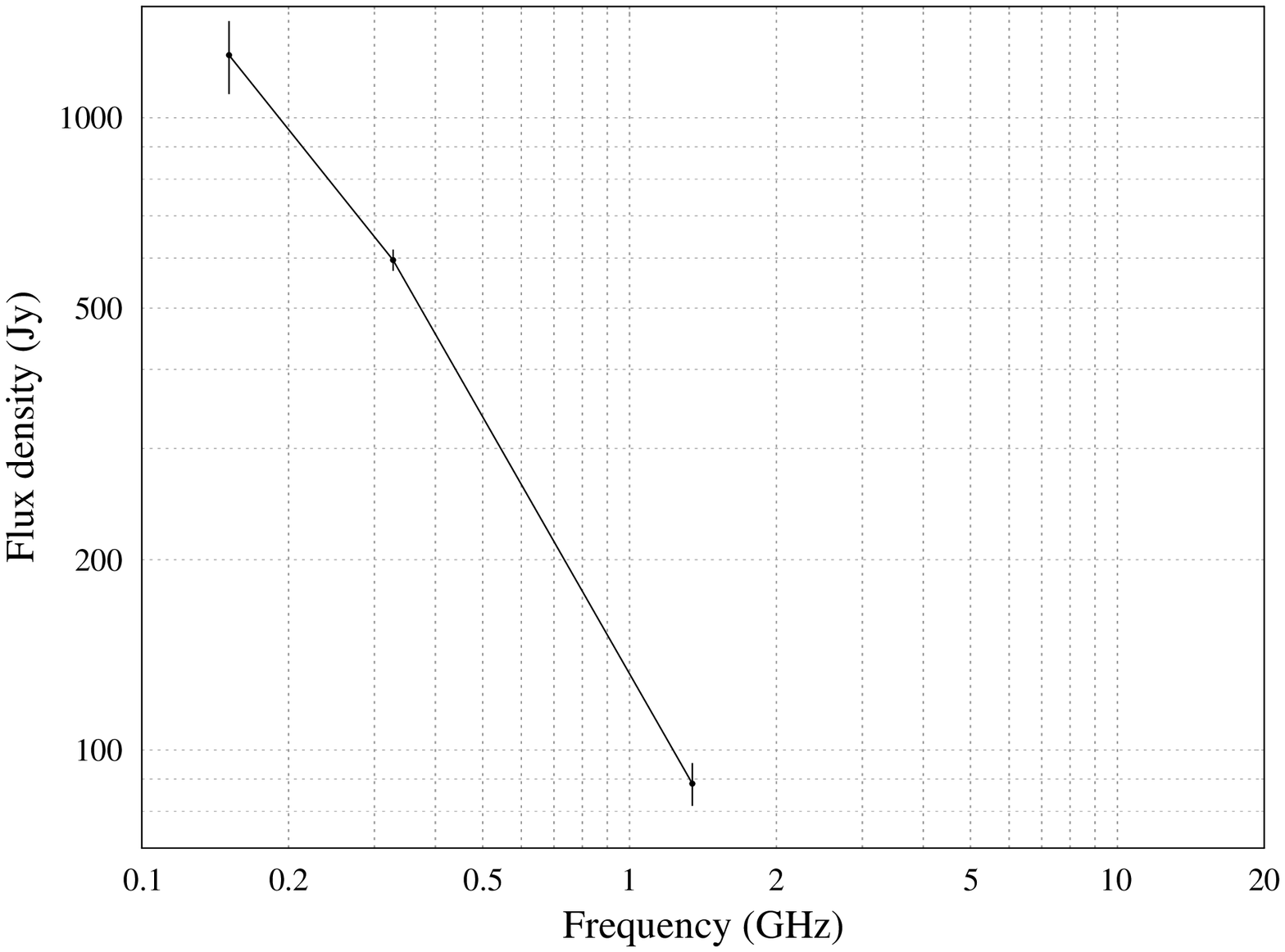}
\end{minipage}
\caption{{\it Upper panel:} The radio spectra for the lobe,
  counterlobe, the bright and weak  hotspots. For details of
  how these flux densities were determined, see text. {\it Lower
  panel:} The difference spectrum between the counterlobe and lobe
  with error bars. \label{fig:radio_spec}}
\end{center}
\end{figure}

At frequencies of 1345~MHz and above the jet becomes quite bright, and
was therefore also subtracted where appropriate. For the 8 and 15~GHz
data we subtract the contribution of the jet from the total lobe flux
density measurement, as the jet is easily separated from the lobe. At
lower frequencies, where the jet is not as easy to disentangle, and as
with the undetected hotspots, we make no attempt to correct for any 
contribution of the jet. However, as the jet at those frequencies is
considerably weaker than the lobe emission, this should not adversely
affect our results. The resulting flux spectra for the lobe and
counterlobe (minus the hotspot and jet contributions), and the
hotspots, are shown in Fig.~\ref{fig:radio_spec}, and the hotspot, jet
and total flux densities are given in Table~\ref{tab:lobe_flux}. 

\begin{table*}
\caption{Measured flux densities in Jy for different components of Cygnus A across six frequencies (in MHz), includeing total flux density values derived from the Baars et al. (1977) scale for comparison, and the percentage of this value recovered by the radio observations for each frequency. Background-subtracted flux density measurements for the hotspots are presented. Hotspots in the lobe are
denoted L1 and L2 for the brighter and weaker respectively. Similarly
for the counterlobe hotspots which are denoted by C1 and C2. The total
(combined) hotspot flux densities for the lobe and counterlobe are given
under L total and C total. In all cases L represents the lobe and C
the counterlobe. Note that this value does not include any flux
from the core for frequencies of 5~GHz and above. The quoted errors
are from Carilli et al. (1991) with the exception of the 151 MHz
value which is quoted from Leahy et al. (1989). The values, in
kpc$^{3}$, from our volume calculations are given in the last 2 columns.}
\begin{tabular}{@{\extracolsep{-2.5mm}}lcccccccccccccccc} 
\hline
	   &               \multicolumn{6}{c}{Hotspots}            &  \multicolumn{2}{c}{Jets} & \multicolumn{2}{c}{Lobes} & Total   & Error &Baars   & \%Baars & \multicolumn{2}{c}{Volume}\\ 
$\nu$      &  L1     &  L2    &  C1     &  C2  & L total & C total & L            & C          & L           & C           &         & (mJy/beam)&    &         & L     & C     \\ \hline
151.0      & 115.7  & 63.7  & 150.4  & --   & 179.4  & 150.4  & --
& --         & 4746        & 5963        & 10709   & 800 & 10841.3 & 98.8  & 68839 & 61257 \\
327.5      & 176.4  & 65.9  & 156.4  & --   & 242.3  & 156.4  &  --
& --         & 2752.7 & 3262.6     & 6015.3 & 230  & 6063.4 &  99.1  & 79790 & 58402 \\
1345.0     & 69.3   & 20.5  & 83.7   & --  & 89.8   & 83.7   &  36  &
47         & 749.8       & 832.2       & 1582.0  & 40   & 1653.9 &  95.7  & 47794 & 39103 \\
4525.0     & 45.2   & 5.7   &  52.4   &  1.8 & 50.9   &  54.2   &  8.7
& 10.8       & 189.4 & 193.7       & 383.1   & 20   & 410.6  & 93.3  & 23562 & 18863 \\
8514.9     &  20.8   &  3.7   &  33.5   &  0.5 & 24.5   & 34.0   & 0.6
& $<$0.1    & 83.4        & 90.5        & 173.9   & 15   & 187.0 & 93.0  & 10447 & 10308    \\
14650.0    & 13.4   &  2.2   &  21.9   &  0.2 & 15.6   & 22.1   &
$<$0.2     & $<$0.2    & 40.5        & 43.9        & 84.4    & 25   & 95.2   &  88.6  & 6462  & 7517    \\ \hline
\end{tabular}
\label{tab:lobe_flux}
\end{table*}

In the case of the 151~MHz MERLIN image there is apparently significant
brightness variation across the lobe, most likely due to residual
sidelobe structure from the brightest points, and the lobe background makes a
significant (and sometimes dominant) contribution to the total flux
density at the location of the hotspots.  For these reasons there will
be larger uncertainties associated with our measured values in this
regime. Consequently, and with reference to Fig.~\ref{fig:radio_spec},
we suspect that the observed non-monotonic radio spectrum of the
strong counterlobe hotspot between 327 and 151~MHz may not be real.

A very interesting result is that the counterlobe is brighter than the
lobe particularly at frequencies below 1345~MHz. The stronger hotspot
is brighter above 5~GHz in the counterlobe than in the lobe. The weaker
hotspot is always brighter on the lobe side. This is easily seen in
the separation in flux density of like-for-like features in
Fig.~\ref{fig:radio_spec}. 

\begin{figure}
\begin{center}
 \resizebox{\hsize}{!}{\includegraphics[angle=0]{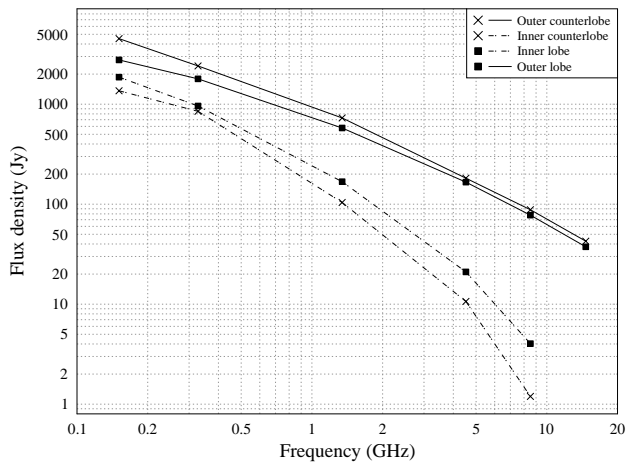}}
\caption{The radio spectra for the inner and outer regions of both
  lobe and counterlobe. From top to bottom: outer counterlobe, outer
  lobe, inner lobe and inner counterlobe. \label{fig:fluxes2}}
\end{center}
\end{figure}

The emission at 327 and 1345~MHz correlates well with the 151~MHz
emission, and exhibits no large-scale differences. To further study
the difference in luminosity of the lobe and counterlobe we plotted radio
spectra for the outer half of the (counter)lobe and the inner half of
the (counter)lobe. Fig.~\ref{fig:fluxes2} shows the spectra for these regions.
The general trend for the spectra to get steeper with increasing
frequency holds true for both the inner and outer regions of the lobe and
counterlobe. This trend is also observed in
Fig.~\ref{fig:radio_spec}. The inner lobes are not detected above the
background noise in the 15~GHz data, and therefore no values are
specified. Fig.~\ref{fig:fluxes2} shows, as expected, that the inner
lobe has a steeper spectrum at higher frequencies than the outer
lobes. At lower frequencies, as Fig.~\ref{fig:fluxes2} shows, the
spectral indices (the slope of the lines) are very similar for both
the inner and outer regions of the lobe and counterlobe.  There is
however a brightness difference manifested in the vertical
offsets. Note also that the  outer counterlobe is brighter than the
outer lobe, see also Table~\ref{fig:fluxes2}, but that the inner lobe
is brighter at all frequencies than the inner counterlobe. 

\section{Discussion}

As is clear from Fig.~\ref{fig:fluxes2} the outer counterlobe is
brighter than the outer lobe at all frequencies. We should note
however that the hotspot-subtracted total lobe flux densities are very
similar for 5, 8 and 15~GHz. We make the assumption that intrinsically
the current lobe and counterlobe should be equally bright at all
frequencies in the rest frame. We note that the light travel time is
small compared to the synchrotron and inverse-Compton loss timescale
especially for low Lorentz factor electrons. We now discuss several
possible explanations for the observed luminosity difference.

\subsection{Absorption}

A possible explanation for the excess emission in the counterlobe at
low frequencies is that the lobe is more heavily absorbed by (fully or
partially) ionised gas. This possibility was acknowledged by
\cite{kassim96} to explain the low-frequency curvature.
\cite{lazio06} used the same explanation for the brightness asymmetry
in the 74~MHz observations, i.e. that the counterlobe is brighter than
the lobe.  However, upon detection at earth, the photons from the
counterlobe have traversed a larger distance and thus passed through
more cluster gas than those from the lobe, which tends to counter the
absorption hypothesis of \cite{kassim96}. There are two possible explanations for an
asymmetry in absorption. Firstly, the column density or temperature
through the cluster in which Cygnus A is situated might vary.  The
Cygnus~A cluster is merging or has recently merged
\citep{markevitch99} and thus has a temperature profile that shows
small scale structure \citep{belsole07}.  In general the counterlobe
side has a lower temperature than the lobe side, as the merger appears
to have occurred on the lobe side. Absorption is more efficient for
lower temperature gas, thus arguing against this
possibility. Alternatively, the column density along the Cygnus arm of
our Galaxy, behind which Cygnus A is located, could vary. In that case
one can assume that the temperature of the gas is much lower and of
order 10$^4$ K.

We tested the absorption hypothesis using two methods. First we
determined the total neutral hydrogen column density assuming solar
abundances \citep{grevesse89} for the two brightest opposing hotspots
in the X-ray image. The measured column density includes any low
ionisation gas. We chose the hotspots as their spectra in this wavelength
range should be accurately modelled by a power-law. Also, due to the
small physical scale of the hotspots contamination from thermal
cluster gas is negligible. 

It is therefore straightforward to measure the absorbing hydrogen
column density. For the less bright and more extended emission, there
is the possibility of different thermal emission components,
complicating the spectral analysis. Although the hotspots appear very
bright on the combined 200~ks {\it Chandra} ACIS-I image, the total
number of counts at their positions is still rather limited. The
hydrogen column density measured for both hotspots is therefore
relatively uncertain. For the western hotspot (i.e. the hotspot in the
lobe) we determined a total hydrogen column density of (3.6 $\pm$
0.4)$\times$10$^{25}$ m$^{-2}$. For the eastern hotspot (i.e. the
hotspot in the counterlobe) the value is (3.5 $\pm$
0.65)$\times$10$^{25}$ m$^{-2}$. \cite{dickey90} quote a hydrogen
column density of 3.5 $\times$ 10$^{25}$ m$^{-2}$ for Cygnus~A, thus
both our measured values are consistent with their value. 

To ensure we were not missing any low ionisation gas, we fixed the
neutral hydrogen column density to the value quoted by
\cite{dickey90}, and added an absorption component assuming a
temperature of 1~eV (1.16$\times$10$^4$ K). In the following analysis
we make the assumption that the neutral Galactic column density does
not vary over the angular size of Cygnus A. We derive 3 $\sigma$ rms
upper limits of 0.6 and 1.7 $\times$ 10$^{23}$ m$^{-2}$ for the
ionised hydrogen column density, $N_{\rm H}$, for the counterlobe and
lobe hotspots respectively. Both values are consistent with no
significant change in column density, either neutral or low
ionisation, at the outer extremities of Cygnus~A. To calculate whether
these upper limits are consistent with the absorption required to
explain the low radio frequency bright excess in the counterlobe, we
use the values assumed by \cite{carilli89} for the electron
density, $n_e$ = 3 $\times$ 10$^5$ m$^{-3}$ (in their paper they quote
this as an upper limit and under the assumption that the absorber
is uniform) and size $d = 3$~kpc for the absorber. From these values and
$N_{\rm H}$= $n_e \times d$ we then calculate a hydrogen column density of
2.7 $\times$ 10$^{25}$ m$^{-2}$, much larger than either measured upper limit
of 0.6 and 1.7 $\times$ 10$^{23}$ m$^{-2}$. 

Another test is to measure the free-free absorption using the spectral
indices and intensities of the different radio frequency images. We
followed the formalism used by \cite{walker00}, using their equations
1 and 2. We assume that the counterlobe is unabsorbed, in order to
derive an intrinsic intensity. We further assume that the intrinsic
intensity is the same for both lobes, an assumption that is based on
the fact that the counterlobe to lobe luminosity ratio decreases to
about 1 by 5~GHz. Furthermore we assume that the intrinsic spectral
index is the same for both lobes. Using the intrinsic intensity we can
then find the extinction parameter $\kappa$ from data at two different
frequencies, which is inversely proportional to frequency squared. The
other variables determining $\kappa$, are temperature, density, and
the path length of the absorber, which should not vary with frequency
for any given point in the lobe. Therefore, finding $\kappa$ of 0.64
for the 151$-$327~MHz and the 5$-$8~GHz data pairs, we can rule out
that free-free absorption explains the counterlobe/lobe luminosity
difference. The other pairs of $\kappa$ are all too similar to allow
for a frequency-squared dependence.

\begin{figure}
\begin{center}
 \resizebox{\hsize}{!}{\includegraphics[angle=0]{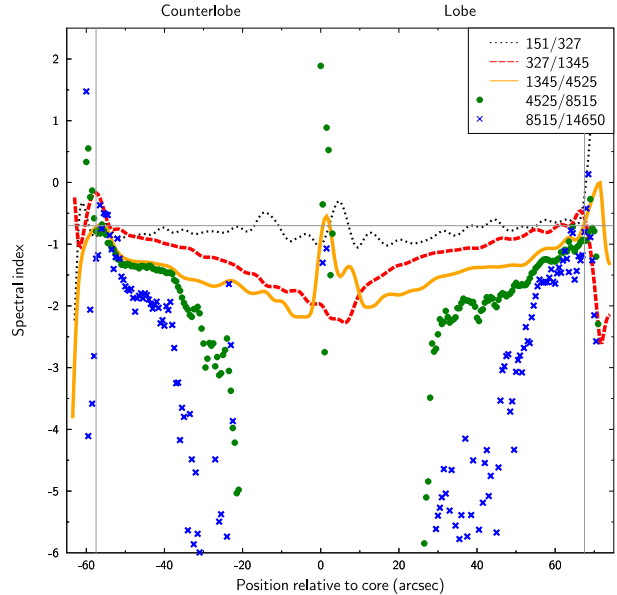}}
\caption{The spectral index for different frequency pairs versus the
  offset from the nucleus in arcseconds. The faint vertical lines indicate
  the location of the inner hotspots. The horizontal line is there to
  indicate that the outer counterlobe spectral index falls off less
  rapidly with distance from the hotspot than for the outer lobe. \label{fig:spectral-index}}
\end{center}
\end{figure}

We should note that the effects of free-free absorption would be more
significant at lower frequencies if this effect was occurring. Thus
the difference in spectral indices should be largest for low
frequencies. Rather, the measured indices at lower frequencies are
very similar for the lobe and counterlobe
(Fig.~\ref{fig:spectral-index}). It seems unreasonable to assume that
the spectra are intrinsically different, but appear to have the same
spectral index due to absorption. This conclusion still holds if just
the counterlobe itself is absorbed.

As an alternative test, assuming that the temperature, metallicity and
Gaunt factor of the absorber are the same for the lobe and counterlobe
side, and that $n_in_e = 2n_e^2$, with $n_i$ the ion density; we can calculate
the ratio of electron densities in the absorber. Using equations 5.16 and 
5.19b from \cite{rybicki86} and rearranging we get
\begin{equation}
j_{\rm cl}/j_{\rm  l} = n_{e, \rm cl}^2/n_{e, \rm l}^2,
\end{equation}
where the subscript cl stands for counterlobe side and l for lobe
side. Substituting the flux densities as those measured from the
151~MHz counterlobe and lobe, we find an electron density ratio of
1.26. The hydrogen column density equation above, which can be
rewritten as $|\Delta H_{\rm H}|$ = 0.12 $n_{e, \rm cl} d$, which yields 3.2
$\times$ 10$^{24}$ m$^{-2}$ for the values of electron density and
size of the absorber assumed by \cite{carilli89}. This is
significantly more than the measured difference in hydrogen column
density of 1.1 $\times$ 10$^{23}$ m$^{-2}$. From the above three
different methods it can thus be concluded that absorption cannot
explain the excess emission observed in the counterlobe at low radio
frequencies, contrary to the conclusion by \cite{kassim96}.

The deduction above is consistent with the fact that the spectral indices
we measure (see Fig.~\ref{fig:spectral-index}) which were made using
matched-resolution image pairs, are nearly symmetric around the core,
which would not be expected if there was significant absorption. 
The spectral indices of the absorbed and unabsorbed lobe of 3C84
studied by \cite{walker00} are very different: $-$0.7 for the
unabsorbed lobe and on average 2.3 in the absorbed lobe. The
absorption is expected to be smaller in Cygnus~A as we do not observe
as large differences in luminosity above 5~GHz, contrary to
observations of 3C84 in which absorption is clearly taking place.

\subsection{Adiabatic expansion \label{sect:adiabatic}}

\begin{figure}
\begin{center}
\resizebox{\hsize}{!}{\includegraphics[angle=0]{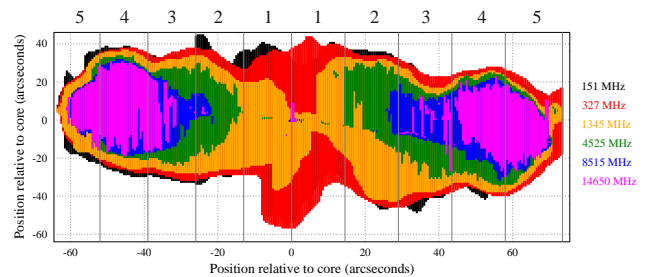}}
\caption{The extent of our assumed emission regions for the different
  radio frequencies: 151~MHz (black, 0.825~Jy/beam), 327~MHz (orange,
  0.207~Jy/beam), 1345~MHz (yellow, 0.1635~Jy/beam), 5~GHz (green,
  2.3~mJy/beam), 8~GHz (blue, 1.2~mJy/beam) and 15~GHz (pink,
  2~mJy/beam). This figure highlights how the extent of the lobes
  increases drastically with decreasing frequency.  The numbered
  regions show how we divide up the lobes to compare like-for-like
  regions. See Section 4.2 for details. \label{fig:volume}}
\end{center}
\end{figure}

Another possible explanation for the luminosity excess at lower
frequencies is that the counterlobe has a slower expansion rate than
that of the lobe. Due to expansion the electrons lose energy and the
magnetic field strength decreases. Any break frequency of the
synchrotron emission moves to lower frequencies, and the total amount
of emission decreases for all frequencies. The effect this has on
log-log spectra (such as Figs.~\ref{fig:radio_spec} and
\ref{fig:fluxes2}) is to shift the curve downwards and to the left
\citep{scheuer68}. Thus at any given frequency, if the spectrum has
concave curvature adiabatic expansion causes the spectrum of the lobe
to appear steeper. Considering the spectrum of the outer lobes, the
two curves have a slightly different shape; for high frequencies the
spectrum of the lobe is certainly not steeper than that of the
counterlobe. At low frequencies the counterlobe has a steeper spectrum
than the lobe. Thus adiabatic expansion of the lobe seems an unlikely
explanation for the difference in brightness.

\begin{figure}
\begin{center}
 \resizebox{\hsize}{!}{\includegraphics[angle=0]{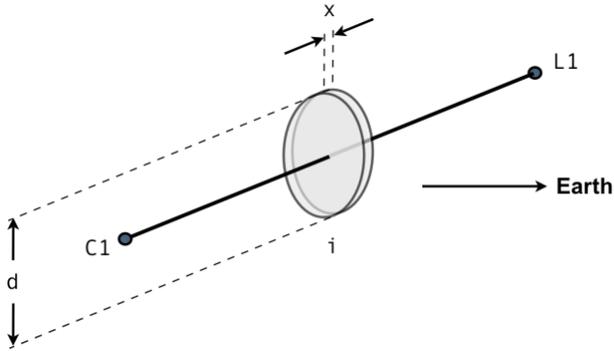}}
\caption{Schematic diagram showing a volume element. Please refer to
  Section 4.2 for details.\label{fig:volume-elements}}
\end{center}
\end{figure}

To calculate the volumes of the lobes we rotate the radio images such that 
the line joining the brightest hotspots are aligned with the
horizontal axis. Each pixel along this axis then represents the length
$x$ of a volume element $i$ which we assume to be a flat cylinder (or
disc), as shown in Fig.~\ref{fig:volume-elements}. The diameter of the element 
is the transverse extent of the radio emission within the threshold
$d$, as defined in Section 3 and shown in Fig.~\ref{fig:volume}. The summation:

\begin{equation}
V = \sum_{i}\frac{\pi}{4}d_{i}^{2} \cdot x
\end{equation}
over all volume elements either to the east or the west of the core
then gives the total volume for the counterlobe or the lobe. 

Table~\ref{tab:lobe_flux} shows that the counterlobe volume is smaller
than that of the lobe with the exception of 15~GHz. Note that our
volume calculation makes a simplistic assumption about the cylindrical symmetry of
the radio emitting region, however we feel this is reasonable in the
absence of any insight into the extent of Cygnus A in the unobservable
third dimension.

A further test is to calculate the expected luminosity ratio
between the counterlobe and lobe, assuming adiabatic expansion and a
tangled magnetic field (despite the indications mentioned above that
the different spectral indices do not favour this model). To do so we
determine the volumes of both regions of interest and the spectral
index. Following the formalism described by \cite{leahy91}, the
fractional luminosity increase due to adiabatic expansion from a state
1 (lobe) to a state 2 (counterlobe) is: 

\begin{equation}
\frac{L_{C}}{L_{L}} = \left(\frac{V_{C}}{V_{L}}\right)^{\frac{-2+4\alpha}{3}}
\end{equation}
where $V$ is the volume, 
and assuming that each lobe starts from the same initial volume and
spectral properties.

As can be seen in Fig.~\ref{fig:volume} we have divided the lobes into
five regions for easy reference. The low frequency excess is contained
within regions 3, 4 and 5 in the counterlobe so we determine the
luminosity ratio with the total volumes of these three regions
($V'_{\rm C}$=30564, $V'_{\rm L}$=32111 kpc$^3$) calculated as per the
method described above.  

Throughout this paper spectral indices are determined for a given frequency by
pairing it with the next highest frequency (i.e. 151/327, 327/1345,
1345/4525). Assuming an average spectral index at 151~MHz across the
source of $-$0.94, the relationship above predicts a brightness excess
in regions 3, 4 and 5 of the counterlobe of 8.0\%, substantially less
than the $\sim$35\% excess that is actually measured in these regions
($L'_{\rm C}$=4801, $L'_{\rm L}$=3412 Jy for 151~MHz). Doing the
calculation for the 5~GHz data (with a spectral index of $-$1.54,
$V'_{\rm C}$=15943, $V'_{\rm L}$=19129 kpc$^3$) the same adiabatic
expansion would predict that the counterlobe should be $\sim$64\%
brighter than the lobe, compared to the 3\% we measure ($L'_{\rm
  C}$=187, $L'_{\rm L}$=181 Jy). 

\begin{figure*}
\begin{center}
 \resizebox{\hsize}{!}{\includegraphics[angle=0]{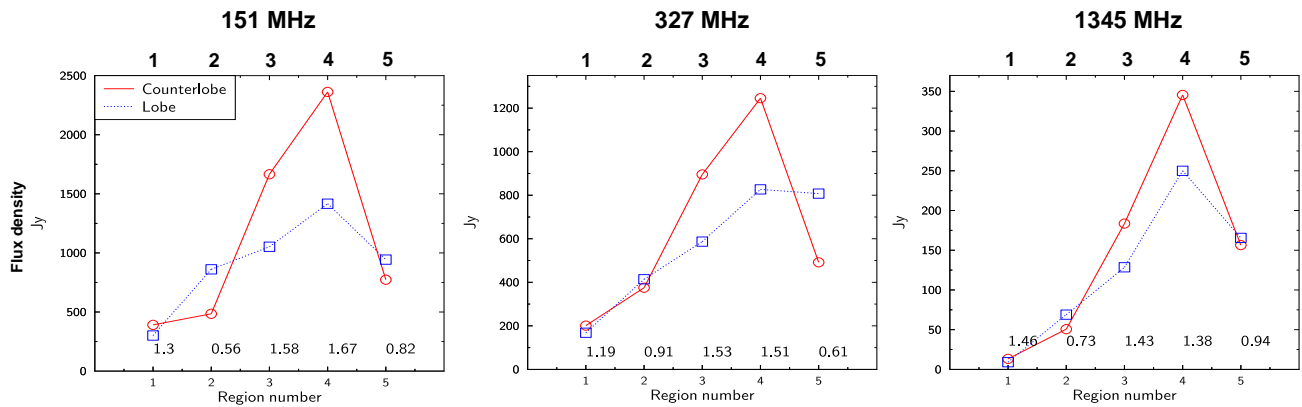}}
\caption{The flux density of five different regions 
  (see text for details) of the counterlobe (solid line, circles) and the lobe
  (dotted line, squares), at 151, 327 and 1345~MHz. Region 1 is the
  region closest to the nucleus, 5 the region which includes the
  hotspots. \label{fig:stripes}}
\end{center}
\end{figure*}

To illustrate the brightness excess in the counterlobe,
Fig.~\ref{fig:stripes} shows the flux densities at 151, 327 and
1345~MHz for the five regions of the lobe and counterlobe. The numbers
towards the bottom of each panel are the counterlobe to lobe ratios of each
plotted value.  Although the flux densities for the lobe and
counterlobe are similar near the nucleus, for the lower two
frequencies the counterlobe is much brighter in the central region of the
lobe, becoming less bright again at the outermost slice, which
contains the hotspots. The difference in flux density between the
counterlobe and lobe is less pronounced for 1345~MHz (and further
diminishes for higher frequencies as seen in Fig.~\ref{fig:radio_spec}).

\subsection{Adiabatic expansion and absorption}

Although neither absorption nor adiabatic expansion alone can explain
the excess flux at low frequencies in the counterlobe, we now consider
whether the combination of both could. The reason why absorption alone
does not work is that the absorption coefficients, derived for the
different frequencies, are very similar, instead of having an inverse
dependence on the square of the frequency. For absorption to be the
correct explanation for the excess brightness we require an increasing
difference in the ratio between counterlobe and lobe flux densities as
we go to lower frequencies. Above we calculated that 8\% of the flux
difference can be explained by adiabatic expansion. We thus
recalculated the absorption coefficients, $\kappa$ assuming that 8\%
of the flux difference is due to adiabatic expansion. For the
151$-$327~MHz pair $\kappa$ = 0.54 and for the 5$-$8~GHz pair
$\kappa$= 0.46, assuming that 64\% of the counterlobe flux at 5~GHz is
explained by adiabatic expansion, as calculated above. Again the
difference between the values of $\kappa$ is too small to allow for
the frequency square dependence.  Therefore we can rule out that a
combination of absorption and adiabatic expansion explains the excess
counterlobe emission.

\subsection{Emission from aged relativistic plasma \label{sect:4.4}}

The excess counterlobe versus lobe low frequency emission can be explained by
an aged relativistic plasma. Considering that in the X-ray image we
detect emission from a relic counterjet \citep{steenbrugge08b},
i.e. the counterjet of a previous episode of jet activity, a possible
origin for this excess radio emission is a relic counterlobe that is
still emitting at low frequencies. The reason that we would only
detect the relic counterlobe and not the relic lobe is the same as the
reason why we only significantly detect the relic counterjet. We look
further back in time when looking at the counterlobe than we do when
looking at the lobe, and thus see less evolved (hence less faded)
plasma. \cite{steenbrugge08b} in their fig. 4 indicate the lightcurve
expected of a jet or lobe that has stopped receiving newly accelerated
electrons.

From Fig.~\ref{fig:stripes} it is clear that near the nucleus the
fluxes measured are very similar, consistent with the idea that near
the nucleus the difference in the look-back times will be very small,
and so the particles should have very similar ages and will have cooled
by roughly the same amount. As expected, the flux density ratio
between counterlobe and lobe emission generally increases with
distance from the nucleus. This increase in ratio is expected from the
look-back time difference, as the difference in travel time becomes larger as
one moves towards the hotspots.

Is there other possible evidence that the excess emission comes from
aged relativistic plasma? At 151~MHz the width of the
brightest part of the counterlobe, which is near the hotspots, and
where most of the excess emission occurs, is significantly greater
than that of the corresponding region in the lobe (see Fig.~\ref{fig:151} where
we optimised the transfer function to highlight the bright
emission). Near the nucleus however, the lobe is wider than the
counterlobe. These width differences have two possible explanations,
namely due to much faster adiabatic expansion of the current
counterlobe compared to the current lobe, or due to an expanded 
aged plasma.

In Section~\ref{sect:adiabatic} we showed that adiabatic expansion
differences cannot explain the excess emission. Therefore we make the
claim that the extra width in the counterlobe is consistent with it
being due to emission from an aged plasma that has expanded. We should
re-iterate that we are assuming that only the excess low frequency
counterlobe emission is due to an aged plasma. However, we cannot rule
out that some of the remaining low frequency emission in both lobes is
due to emission from an aged plasma.

\begin{figure*}
\begin{center}
 \resizebox{\hsize}{!}{\includegraphics[angle=0]{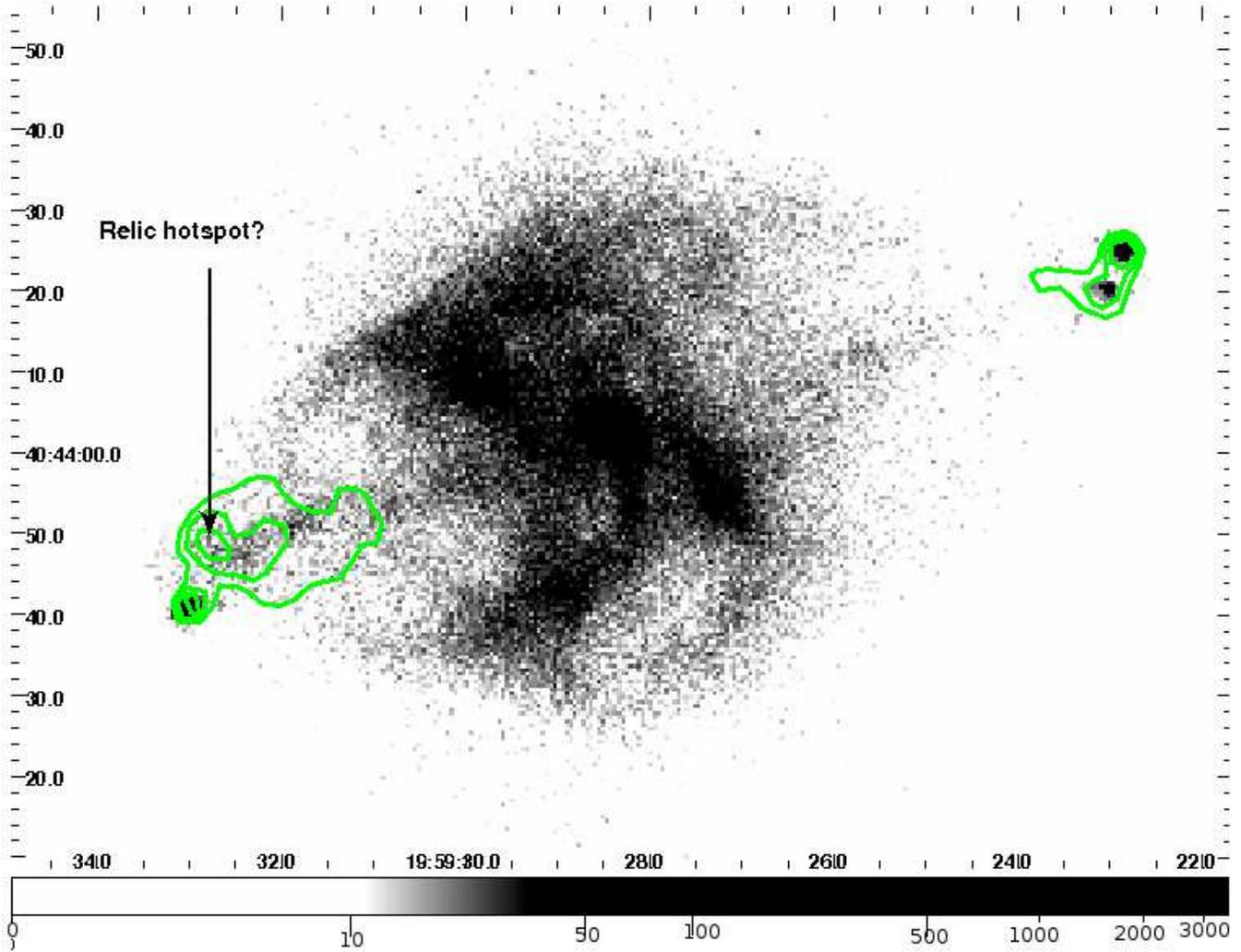}}
\caption{The 0.2$-$10 keV X-ray image of Cygnus~A optimised in
  transfer function (grey scale indicates the number of counts per
  pixel and is logarithmic) to see the emission co-spatial with the radio
  lobes. The 151~MHz high brightness emission regions are overlayed as
  green contours, with contours at 75, 97.2, 119.4, 141.6, 163.8 and
  186 Jy/beam. There is excess 151~MHz emission where the X-ray relic
  counterjet terminates, which is a possible location for the relic
  hotspot, is indicated.  Note that the X-ray emission brightens at
  the outer contour plotted in the northern part of the
  lobe. \label{fig:relic}}
\end{center}
\end{figure*}


Considering the close spatial association of the relic X-ray
counterjet with the aged plasma observed at 151~MHz, we identify the
aged plasma with a part of the relic counterlobe that was formed by
the relic counterjet. At the extreme end of the relic X-ray counterjet
there is a region of perpendicular X-ray emission which is quite
separate from the current hotspots. This is a possible location for
the corresponding relic hotspot. The emission mechanism here could not
possibly be synchrotron but it could be explained as cosmic
  microwave background photons inverse-Compton scattering off the electrons
(ICCMB) present in the plasma which once formed the
hotspot. Fig.~\ref{fig:relic} highlights the location of this possible
relic hotspot. This region is the brightest point on the 151~MHz image
(other than the current hotspots), however, in X-rays there is an
equally bright region between the 2 outer contours plotted in Fig.~\ref{fig:relic}. Within the
contour labelled relic hotspot there are 395 counts above the
background, while the flux density is 149 Jy/beam with a beam size of
3$^{\prime\prime}$ by 3$^{\prime\prime}$.

\subsubsection{Compression of relic counterlobe}

The lifetime of the low frequency synchrotron-emitting electrons
should be shorter than that of those scattering off the CMB to produce
X-rays. If we take synchrotron and inverse-Compton losses and losses
due to adiabatic expansion into account, as well as the decrease of
the magnetic field strength over time, and the light travel-time
effects then we cannot adequately explain why we do detect the
relic counterjet in X-rays, yet still see the low frequency radio
emission from the counterlobe.

To explain why we detect a relic counterlobe and a possible relic
counterhotspot, we suggest that the current counterlobe is compressing
the relic counterlobe plasma and in doing so increasing its magnetic
field strength and re-energising the electrons (the frequency of
emission being proportional to the magnetic field strength).  The
current counterjet lies to the south of the relic counterjet, and
therefore one would expect more radio emission to the north of the
current counterjet, as is observed (see Fig.~\ref{fig:15GHz}). This
implies that a part of the relic counterlobe is embedded with the
current counterlobe, or is close enough so as to be compressed by
it. We should also note that the relic counterlobe is likely much
larger than the portion that is currently visible, as we are only
detecting the plasma in the relic lobe that is re-energised.

Fig.~\ref{fig:relic} shows one more interesting feature: on the
northern side the X-ray brightens just outside the outer contour
plotted (at about 19$^h$59$^m$32$^s$ in Right Ascension and 40$^{\circ}$43$^{\prime}$54$^{\prime\prime}$ in Declination). A possible explanation is that the relic counterlobe as it
expands is coming into contact with cluster gas and either causing a
shock wave resulting in heating and brightening of the cluster gas, or
the expansion is subsonic and therefore causing compression of the
intracluster medium. The denser gas will cool faster than its
surroundings and therefore appear brighter. Data of much higher
quality are needed to disentangle these possibilities.

\subsection{Morphology}

Higher frequency images of Cygnus~A (see fig. 7 in
\citealt{steenbrugge08a}), show that there is a corresponding lack of
radio emission at the location of the relic X-ray counterjet near the
hotspot in the counterlobe. However, at 151~MHz there is excess
emission overlaying the relic counterjet (see
Fig.~\ref{fig:relic}). The correspondence between 0.2$-$10~keV X-ray
and 151~MHz emission for the counterlobe is, with the exception of the
excessive transverse extent of the outermost contour plotted, rather
good. This suggests that part of the 151~MHz emission originates in
the relic counterjet.

We identify the extent of the 151-MHz emitting aged relativistic
plasma coinciding with the outer contour on the counterlobe side in
Fig.~\ref{fig:relic}. The magnetic field in the relic hotspot inside
the relic counterlobe is likely to still be higher than that of the
surrounding area. Thus lower Lorentz factor electrons can still emit
at observable but low frequency radio bands. This is a possible
explanation as to why the likely relic hotspot is the brightest part
of the excess emission in the counterlobe compared to the lobe (if we
neglect the current hotspots). At even lower Lorentz factors, the
electrons inverse-Compton scattering the cosmic microwave background
will produce X-ray emission at $\sim$ keV energies, enabling the
detection of the relic counterjet. This relic structure probably
extends all the way to the nucleus, and certainly extends much further
towards the nucleus than the excess counterlobe low frequency radio
emission.

\subsubsection{Impact of the relic counterjet on high frequency emission}

\begin{figure}
\begin{center}
 \resizebox{\hsize}{!}{\includegraphics[angle=0]{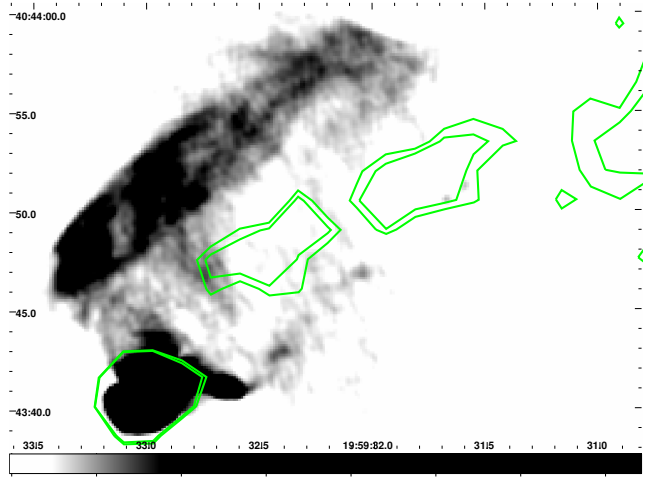}}
\caption{The 15~GHz image of the counterlobe, with the contours
  detailing the X-ray detected relic counterjet at 14 and 15 counts
  and current hotspot. Note the lack of radio emission along the
  trajectory of the relic counterjet which is shown in green
  contours. The grey scale indicates the 15~GHz intensity and is
  in Jy per beam. \label{fig:15GHz}}
\end{center}
\end{figure}

As mentioned above the higher frequency radio data, especially at
15~GHz, show a lack of radio emission overlaying the relic X-ray
counterjet (see Fig.~\ref{fig:15GHz}). This is consistent with the
feature being a true relic counterjet as there are insufficient
high-$\gamma$ particles to cause synchrotron emission; we only see
ICCMB emission at keV energies from $\gamma$$\sim$1000 particles. Currently-active
jets in FR II galaxies tend to be brighter relative to the lobes at
high frequencies which is certainly the case for the current jets in
Cygnus~A. In Fig.~\ref{fig:relic} the current jets are not visible in
the 151~MHz contours plotted or the X-ray image.

\begin{figure}
\begin{center}
 \resizebox{\hsize}{!}{\includegraphics[angle=0]{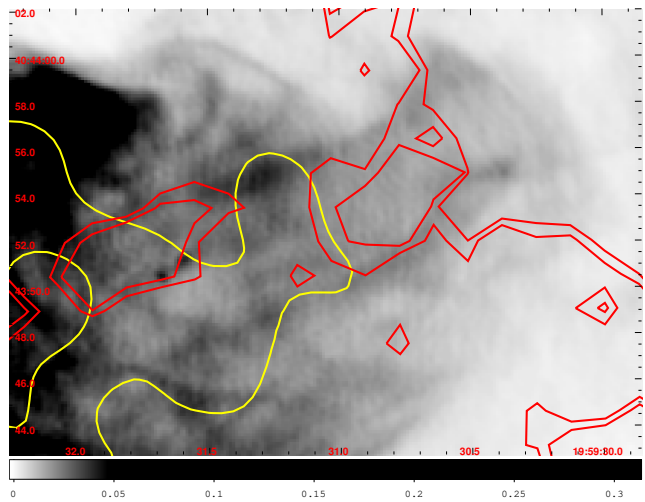}}
\caption{Detail of the inner counterlobe at 5~GHz, showing the
  `antennae', two seemingly parallel, fairly straight, brightened
  features. The red contours are for 14 and 15 X-ray counts in the 0.2
  $-$ 10~keV band. They trace mainly the relic X-ray counterjet and
  the outer edge of the bright thermal gas near to the nucleus. The
  yellow contours are at 78 and 96 Jy per beam in the 151~MHz
  image. Note how the `head' of the outer yellow contour has a similar
  width to the separation between the antennae and lies in between the
  two brighter X-ray contours detailing the relic X-ray
  counterjet. The grey scale denotes the 5~GHz intensity in
  Jy/beam.  \label{fig:5GHz}}
\end{center}
\end{figure}

In the 5~GHz image \citep[see figs.~1 and 4 in][]{steenbrugge08a} at
the inner edge of the counterlobe there is quite a bit of substructure
in the brightness profile. There is a very weak `ring' and there are
two `antennae`, which are two nearly-parallel, bright, narrow, linear
features. In Fig.~\ref{fig:5GHz} we show these features and overlay in
red contours the relic X-ray counterjet and a section of the outer
edge of the bright thermal emission near the nucleus; see
Fig.~\ref{fig:relic} for an X-ray image detailing the relic
counterjet. The inner part of the X-ray relic counterjet lies between
the two antennae. The electrons producing the 5~GHz emission are
interacting with the relic counterjet. The brightening of the antennae
is best explained by the relic counterjet halting the expansion of the
current counterlobe plasma. Due to the slower adiabatic expansion
rate, the losses are smaller, and the magnetic field will be somewhat
higher, thus the electrons will continue to emit at higher
frequencies. Furthermore, the increased density of the electrons will
result in brighter emission.

The western-most part of the 151~MHz contours in
Fig.~\ref{fig:relic} overlays part of the antennae. This part of
the 151~MHz contour lies between the two X-ray counterjet contours in
Fig.~\ref{fig:5GHz}. This 151~MHz contour has about the same width as
the distance between the two antennae.

The 151~MHz contour lying between the X-ray contours in
Fig.~\ref{fig:5GHz} coincides with a northwards shift of the X-ray
counterjet. This occurs just before the location where the current
counterjet deflects by an angle of about 27$^{\circ}$. There are two
possible explanations for this shift.  It could be due to the
precession of the relic counterjet \citep[see Fig.~9
in][]{steenbrugge08a} or it could be the case that the current jet
pushes the relic counterjet plasma northwards before itself deflecting
southwards.

\subsubsection{Did the relic lobe rise?}

Current theories \citep[e.g.][]{churazov01} state that the relic lobes
should buoyantly rise through the intracluster medium, at least as
long as it is in direct contact with the cluster medium. The authors
state a terminal velocity of bubbles which is of order 400 km
s$^{-1}$. The relic lobe is estimated to have been able to rise for 4
$\times$~10$^7$ years (a timescale of the same order as the duration
of current jet activity), which would mean that the lobe has risen by
about 17~kpc. We cannot test this hypothesis as we cannot trace the
X-ray-detected relic counterjet back to the nucleus due to the excess
thermal emission in this region.  Thus it is possible that the relic
counterlobe did rise by 17~kpc, and the close spatial relation between
the end of the relic counterjet and the current hotspots is
coincidence.

\subsubsection{Disentangling ICCMB from relic lobe and thermal cluster emission}
A possible confirmation of relic lobe emission is to look for excess
X-ray emission, because once particle
energies are low enough, the electrons will inverse-Compton scatter
off the cosmic microwave background (ICCMB) and emit X-rays. However
in X-rays, at the site of the lobes further out from the nucleus, the
cluster background contributes a significant fraction. Thus, if the
cluster emission is not centered on the Cygnus~A galaxy, this will
cause a difference in luminosity between the lobes. A further
complication is that the temperature of the cluster gas we fitted in
the vicinity of the lobes is very different for the counterlobe and
lobe \citep[see also][]{belsole07}; e.g. due to a recent or ongoing
cluster merger. A final complication is that the relic counterlobe
observed at 151~MHz might compress and thereby brighten the
intracluster gas. Thus the interpretation of the luminosity over the
area of the ellipses used for fitting the lobes in X-rays is not
straightforward. The values of luminosity per unit area (in 10$^{33}$
Watts per square arcsecond) are 2.45 and 3.35 for the lobe and
counterlobe respectively. Any ICCMB emission should, to first order,
have a spectrum closely resembling that of the Cygnus A synchrotron
spectrum at $\gamma$$\sim$10$^{3}$, however to distinguish between the
multi-temperature thermal cluster gas component and the ICCMB
component, much higher signal-to-noise and higher resolution spectra are needed.

\subsubsection{Duration of previous epoch of jet activity}

The relic counterlobe and relic counterjet partly overlay the outer
part of the current counterlobe. There is no evidence of relic
emission that extends beyond the current hotspot, and the possible
relic counterhotspot is situated well within the current counterlobe.
Relic lobes are supposed to buoyantly rise in the intracluster medium,
and not fall back towards the galaxy nucleus. This indicates that the
previous epoch of jet activity lasted for slightly less time than the
current period of jet activity, assuming the hotspot advance speed to
be the same for both epochs of jet activity. It is possible that the
hotspot advance speed was slower during the previous epoch of jet
activity, as the intracluster medium density was probably higher. The
duration of the current epoch of jet activity is of order 4 $\times$
10$^7$ years, assuming a hotspot advance speed of 0.005$c$
\citep{alexander96}.

\cite{schoenmakers00} constrain the interruption of jet activity for
B~1834+620 to be a few Myrs, very similar to our constraint from the
relic X-ray counterjet of about 10$^6$ yrs
\citep{steenbrugge08b}. \cite{shabala08}, modelling current radio loud
galaxies, derive the mean jet and quiescent lifetimes per galaxy mass
bin. They suggest jet lifetimes which are a bit model dependent, that were just short of the lifetime of
the current jets in Cygnus~A. However, they claim that the quiescent
phase is significantly longer than the timescale derived in this
study, namely between 2 $\times$ 10$^7$ to 10$^8$ years, instead of
10$^6$ years. A possible explanation for the difference in the
quiescent timescale of Cygnus~A and those modelled by \cite{shabala08}
is that Cygnus~A is in a merging cluster of galaxies, which might
alter the accretion properties of the nucleus.

\section{Conclusion}

We have presented a detailed analysis of the low frequency radio
images of Cygnus~A, which show a marked excess emission in the outer
counterlobe compared to the outer lobe. We explain this excess in
terms of emission from a relic counterlobe and relic counterjet. The
relic counterjet detected at 151~MHz traces the outer parts of the
relic counterjet detected in the 200~ks {\it Chandra} image. Current
counterlobe plasma is compressing and re-energising the relic plasma,
and thus making it visible again at low radio frequencies. We
calculate that the duration of the previous epoch of jet activity was
slightly less than that of the current epoch of jet activity, namely
about 10$^7$ years.

\section*{Acknowledgements}
We thank T. Joseph W. Lazio and Paddy Leahy for allowing us to use
their data. The authors would like to thank Paul Goodall for his technical
computer help. The authors would like to thank the referee, Robert Laing for
helpful comments.

\bibliography{references}

\label{lastpage}

\end{document}